\documentclass[aps,prl,twocolumn,showpacs,reprint,superscriptaddress,footinbib]{revtex4-2}
\usepackage{graphicx}
\usepackage{dcolumn}
\usepackage{amssymb}
\usepackage{bm}
\usepackage[utf8]{inputenc}
\usepackage{amsmath,amsfonts,amssymb}
\usepackage{color}
\usepackage[sf]{subfigure}
\usepackage{fouriernc}
\usepackage{xcolor}

\usepackage[colorlinks=true,
            linkcolor=blue,
            urlcolor=blue,
            citecolor=blue,
            unicode,
            pdfencoding=auto]{hyperref}
\usepackage{etoolbox}

\usepackage{soul}

\begin{document}
\bibliographystyle{apsrev4-2}

\title{Graviton Chirality and Topological Order in the Half-filled Landau Level}

\author{F. D. M. Haldane}
\affiliation{Physics Department, Princeton University, Princeton, New Jersey 08544, USA}

\author{E. H. Rezayi}
\affiliation{Physics Department, California State University Los Angeles, Los Angeles, California 90032, USA}

\author{Kun Yang}
\affiliation{NHMFL and Department of Physics, Florida State University, Tallahassee, Florida 32306, USA}


\begin{abstract}

The fractional quantum Hall state at Landau level (LL) filling factor $\nu=5/2$ is extremely interesting because it is likely the first non-Abelian state, but its precise nature remains unclear after decades of study. We demonstrate this can be resolved by studying the chirality of its graviton excitations, using circularly polarized Raman scattering. We discuss the advantage of this bulk probe over the existing edge probes.

\end{abstract}

\date{\today}

\maketitle

{\em Introduction and Motivation} --
Non-Abelian fractional quantum Hall (FQH) liquids are arguably the most exotic quantum states of matter, which can provide a platform for topological quantum computation. The most promising candidate for such a liquid is the one at Landau level (LL) filling factor $\nu=5/2$\cite{Willett1989}, and the leading candidate states (based on extensive numerical studies\cite{Morf1998,Rezayi2000,Wan2006,PhysRevB.77.165316,Moller2008,Peterson2008,PhysRevB.79.115322,HaoWang2009,Wojs2010,Morf2010,Feiguin2008,Storni2011,Pakrouski2015,WZhu2016}) are the Moore-Read (MR) Pfaffian state \cite{Moore1991}, and its particle-hole conjugate partner, the anti-Pfaffian (APf) state \cite{Levin2007,SSLee2007}, both describing electrons in a half-filled LL. In the absence of LL mixing and other symmetry-breaking perturbations, a half-filled LL possesses particle-hole symmetry, as a result of which the MR and APf states are exactly degenerate. LL mixing breaks particle-hole symmetry and appears to favor the APf state\cite{Rezayi_Simon_LLmix,Zaletel2015,Rezayi_2017}.
The situation is much murkier on the experimental front. It has been long believed that the MR and APf states, while topologically distinct,  can only be distinguished in their edge properties. As a result existing experiments attempting to determine the nature of the 5/2 state have been focused on the edge (for a review of earlier experimental work that also includes bulk spin polarization measurements which are consistent with both MR and APf states, see Ref. \cite{Lin_2014}). Among them perhaps the most direct probe is the recent thermal Hall conductance measurement \cite{Banerjee2018}. While the discovery of half-integer quantization definitely points to the non-Abelian nature of the 5/2 state, its specific value turns out to be consistent with neither the MR nor APf state, but suggests a particle-hole symmetric state instead. This (apparent) particle-hole symmetry could be due to the spatial mixture of MR and APf liquids in the sample, that form either spontaneously\cite{Wan2016} or due to disorder that locally breaks the particle-hole symmetry\cite{Mross2018,Chong2018,Lian2018}, which could yield an edge structure that gives rise to the measured thermal Hall conductance. The viability of this scenario is currently under debate\cite{Simon2019,zhu2020topological}. Another controversial explanation of the experiment is the lack of equilibration at the edge\cite{Simon2018b,Feldman2018,Simon2018a}, which is an extrinsic effect. There is, of course, the possibility of an intrinsically particle-hole symmetric FQH state known as PH Pfaffian (PH Pf)\cite{DTSon2015,Feldman2016}, but none of the numerical studies \cite{Mishmash-2018,Mross-2020,yutushui2020largescale,EHR-KP-FDMH} have seen a clear
gapped phase or a state that can energetically compete with either the MR or APf state\cite{EHR-KP-FDMH} (see also\cite{Balram-2018}).

In this paper we point out that the MR, APf, PH Pf (or any other intrinsically particle-hole symmetric FQH state), and in principle their spatial mixtures, can be distinguished by measuring the chirality of a {\em bulk} geometric excitation termed graviton\cite{ChiralGraviton}, which is accessible via polarized Raman scattering\cite{golkar,ChiralGraviton}. In our earlier work\cite{ChiralGraviton} we demonstrated that for electron states (like those in the Laughlin sequence with $\nu=1/m$) the gravitons carry spin -2, and pointed out their particle-hole conjugate states at $1-\nu\ne \nu$ the chirality is reversed and gravitons carry spin +2 (see also Ref. \cite{son2019chiral}). This, however, leaves the situation ambiguous at the particle-hole symmetric filling factor of $\nu=1/2=1-\nu$. It has already been demonstrated\cite{ChiralGraviton} that MR graviton carries spin -2. The APf graviton then must carry spin +2, while both chiralities should be present in a particle-hole symmetric state. Should there be a mixture among these different states, the local chirality can be revealed as long as the probing light can be localized in a region smaller than the domain size. In addition to the obvious and potentially far-reaching experimental relevance, our results also reveal the deep connection between the geometric\cite{Metric,qiu,boyang12,yang,xiluo,bimetric,son2019chiral} and topological\cite{Wen1995} aspects of FQH effect (which has been perhaps somewhat under-appreciated thus far), and point to the possibility of {\em bulk} probes of topological order (for an earlier suggestion in this general direction see Ref. \cite{YangHalperin}).

{\em Models and graviton operators for the 5/2 state} -- As shown in Refs. \cite{yang,ChiralGraviton}, electrons in an LL couples to an external oscillating metric through a set of 2-body graviton operators, whose spectral functions describe the absorption rate of ``gravitational wave" propagating through the system. The graviton operators we employ here are different from their lowest
LL counterparts\cite{ChiralGraviton} and are modified by the presence of a non-trivial LL
form factor, and can be derived the same way as in Ref. \cite{yang}:
\begin{eqnarray}
\hat{O}^{(2)}_\pm(n)&=&\sum_{q_x,q_y}(q_x\pm iq_y)^2 V(q) e^{-q^2/2}\bar{\rho}({\bm q})\bar{\rho}(-{\bm q})F_n(q),\\
F_n(q)&=&\lvert L_n(q^2/2)\rvert^2-2 L_n(q^2/2)L_n^{\prime}(q^2/2),
\end{eqnarray}
where $n$ is the LL index, $V(q)$ is the Fourier transform of the Coulomb potential, $L_n$ is the
$n$th Laguerre polynomial, the projected density operator is
$\bar{\rho}({\bm q})= \sum_n e^{-i{\bm q}\cdot {\bm R_n}}$, and ${\bm R}$ is the
guiding center coordinate. The prime on $L_n$ signifies the derivative with respect
to the argument. The wave vector $q$ is measured in units of inverse magnetic length $1/\ell$, where $\ell=\hbar/eB$. $\hat{O}^{(2)}_\pm(n)$ describe coupling to the ``gravitational wave" with opposite (circular) polarizations, that change the angular momentum of the electron liquid by $\pm 2$ respectively.

The Hamiltonian for the Coulomb repulsion for the $n$th LL is
\begin{eqnarray}
H(n)&=&\frac{1}{2}\sum_{q_x,q_y} V(q) e^{-q^2/2}\bar{\rho}({\bm q})\bar{\rho}(-{\bm q})f_n(q),\\
f_n(q)&=&L_n^2(q^2/2).
\end{eqnarray}
In this work we ignore inter-LL  transitions (or LL-mixing) and focus on the
valence electrons at $\nu=5/2=2+1/2$ that half-fill the second LL with index $n=1$.
The form factors of the
graviton operator  and the Hamiltonian simplify to $F_1(q)=(1-q^2/2)(3-q^2/2)$
and $f_1(q)=(1-q^2/2)^2$ respectively.  
 In some cases we have also
increased the first Haldane pseudopotential of the Coulomb repulsion by a small amount. It is
important for our purposes to
also break particle-hole symmetry by introducing a weak
3-body interaction.  The exact form is immaterial and we choose the simplest case for which the MR
state is a zero energy ground state. This is a repulsive interaction that penalizes the closest
approach of 3-particles\footnote{This is called the 3-body pseudopotential with relative angular
momentum 3\cite{Simon-Rezayi-Cooper-3Bpp}.}.  We will also use its
attractive counterpart by flipping its sign. Such additional pseudopotentials terms also contribute to the graviton operators, in a way that do not involve LL form factors (see Ref. \cite{ChiralGraviton}).

In experiment, LL-mixing breaks PH symmetry by
generating a slew of 3-body pseudopotentials from the 2-body Coulomb repulsion,
which have been calculated
perturbatively\cite{Sodemann_AHM_LLmix,Peterson_Nayak_LLmix} in the LL mixing parameter $\kappa=\varepsilon/\hbar \omega$, where $\varepsilon= e^2/4\pi\epsilon\ell$ is the Coulomb interaction scale, and $\epsilon$ is the dielectric constant of the material.
In most of what follows we quote energies in units of
$\varepsilon$. We also set $\hbar=1$ and ignore the width of the electron layer.
For weak LL-mixing the strongest component corresponds to the MR
pseudopotential and is negative:  -0.0147$\kappa$.

{\em Numerical Calculations-}  Our calculations are on high symmetry  tori,
 namely square and hexagonal geometries. These are somewhat complementary and are
helpful in discerning
finite size effects. Below we review the known characteristics for both MR and APf model states (exact ground states of idealized 3-body model Hamiltonians)
as well as for generic states. For even numbers of electrons
the topological sectors (excluding the 2-fold Center-of-Mass degeneracy) are either a triplet (hexagonal)
with 3-fold point symmetry or split into a doublet and a singlet
for square symmetry. For the model Hamiltonians, all 3 ground states are degenerate with zero
energy
in any
geometry.  Only their respective crystal momenta are different for different geometries.
In  hexagonal geometry these are  at the 3 corners of the Brillouin zone (BZ).
In the case of the square unit cell
the singlet is  at the zone corner (ZC)(1,1),
while the doublet is at the zone boundary (ZB) (0,1)(1,0).
For generic states in the presence of PH symmetry and
for even electrons, the K-vectors of the topological sectors are the same as in
the model states. The degeneracy, however, is different for square geometry.
There is a small splitting of energy between the singlet and the doublet (ZB). Depending on size both the
singlet and the doublet could become the absolute ground state. In our calculations we have
assumed that both are valid candidates irrespective of which one is the absolute ground state.
The splitting is a finite-size effect and the degeneracy is recovered for large sizes.

For the model Hamiltonians with odd numbers of electrons there is one zero energy ground state with
$K=0$ at the zone center, corresponding to the only topological sector for all geometries.

For the generic case, in hexagonal geometry and depending on
whether the number of electrons modulo 6 is one or not, the ground state is a singlet or a doublet
respectively. Both topological sectors of the MR and APf are represented by the doublet\cite{papic_etal2012}.  This is an interesting case and  we will return to discuss it later.

In all cases we calculate the spectral functions of the graviton operators\cite{ChiralGraviton}:
\begin{equation}
I_{\pm}(\omega)=\sum_n \lvert \langle \Psi_0 \lvert \hat{O}^{(2)}_\pm\rvert \Psi_n \rangle \rvert^2 \delta(\omega-\omega_n),
\label{eq:graviton-operator}
\end{equation}
where $\lvert\Psi_0\rangle$ is a
ground state, which is included in the sum over intermediate states. As a result the total graviton
weight can be normalized to one by dividing the RHS of the above by $\langle\Psi_{0\pm}\vert \Psi_{0\pm}\rangle$,
where $\vert\Psi_{0\pm}\rangle=\hat{O}^{(2)}_\pm \lvert \Psi_0\rangle$, so that
$\int  I_{\pm}(\omega)d\omega=1.$

{\em Square Geometry} --
In this geometry for even number of electrons and the ground state doublet there is a conserved unitary
operator that
results from the product of two anti-unitary
 mirror and PH conjugation operators. 
The entire energy spectrum can be classified by a $Z2$ parity quantum number.  However, for
ZB (ground and excited) states  the chiral graviton operator has mixed
parity.  That is, the real and the imaginary parts of $\hat{O}^{(2)}_\pm$ produce states with
opposite parities. This means the two parts are not present simultaneously and
hence the graviton weight is always non-zero.
A finite graviton weight for the ground states, however, is an undesirable effect and
will be removed below.

In contrast, for the singlet ground state as well as the excited states, there are angular momentum
selection rules irrespective of the presence or absence of PH symmetry.
Some states have a finite graviton
weight and some not, according to whether their
angular momentum is within $\pm 2$ of the ground state.
However, when PH symmetry is broken the weights are different for the chiralities $\pm 2$, but they
occur for the same states since on a square the discrete angular momentum $ 2 = -2 \mod(4)$.

Since the energies and the graviton weights are identical for the case of degenerate ground states, we include them in the intermediate sum of Eq. \ref{eq:graviton-operator} and trace over the ground states.
It proves convenient to combine the two ZB ground states as follows:
\begin{equation}
\lvert\Psi_0\rangle_\pm=\frac{\lvert \psi_0\rangle_1\pm\lvert \psi_0\rangle_2}{\sqrt{2}}.
\end{equation}

\begin{figure}[b]
\centering
{\includegraphics[height=2.8in,width=3.6in]{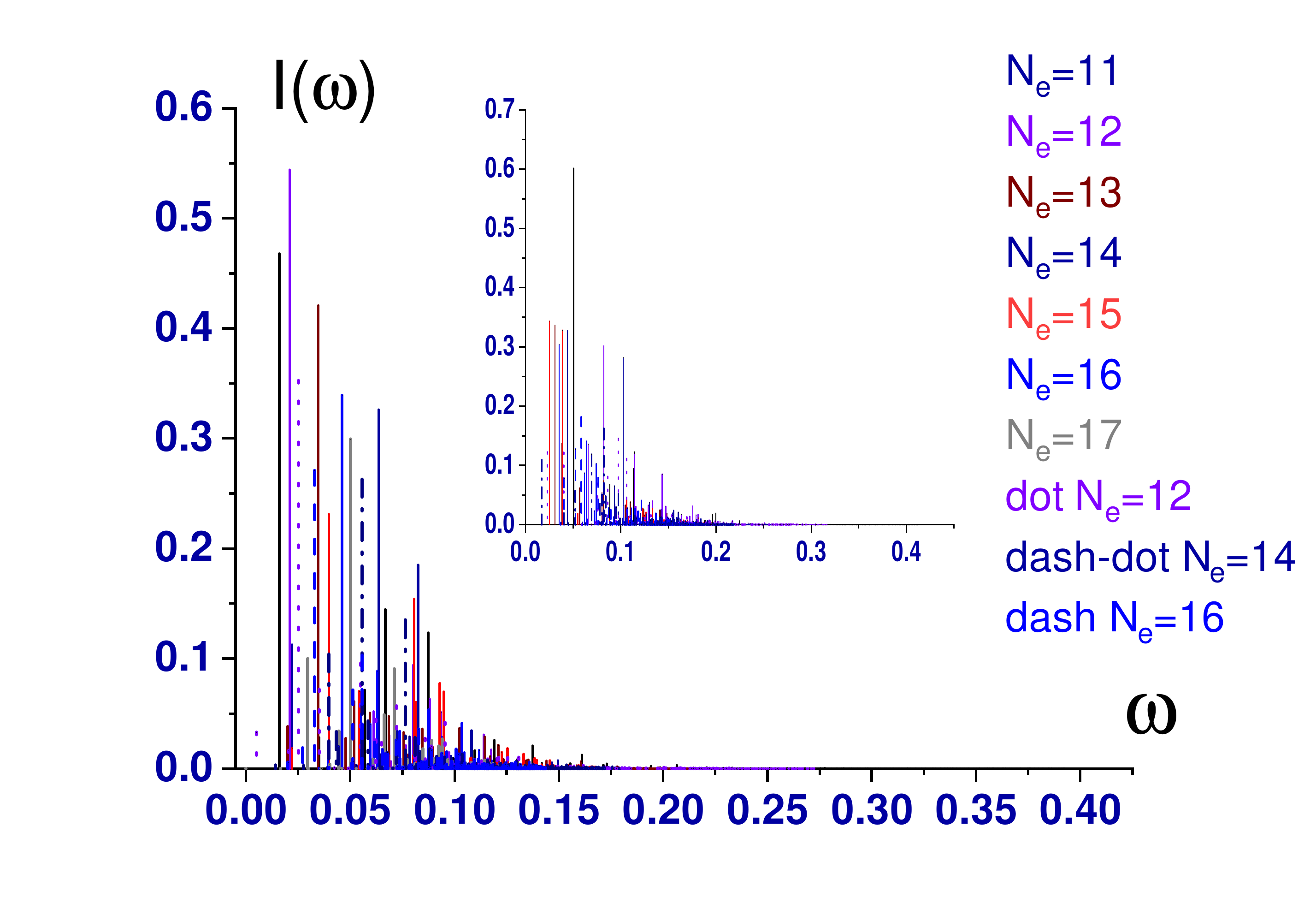}}
\caption{\label{fig:GS_Coul}%
Color online -- Graviton spectral functions for 11-17 electrons on a square
unit cell.  For even number of electrons  we have included data for both ZC and ZB (dotted lines).
discussed in the text. We have only shown the spectrum for positive chirality
(with angular momentum +2). In the inset we have added a $v_1 =0.035$ Haldane pseudopotential \
to the $n=1$ Coulomb interaction. In an isolated n=1 LL the overloaps with MR or APF states are at
or near their maximum for this $v_1$.
The graviton spectrum for negative chirality, by
particle-hole (PH) symmetry, is identical to the one shown as verified.
}
\end{figure}
The wavefunctions in the two (1,2) sectors have different translatonal quantum numbers and are orthogonal. The graviton operator preserves these quantum numbers and hence the matrix elements over
the excited states are now included for both sectors. Because of orthogonality, the intermediate sum
over the excited states separates into two sums. The contribution of the ground state to the sum is
the square of
\begin{equation}
_1\langle\psi_0\lvert\hat{O}^{(2)}_\pm\rvert\psi_0\rangle_1 + _2\langle\psi_0\lvert\hat{O}^{(2)}_\pm\rvert\psi_0\rangle_2=0,
\end{equation}
thus dropping out as verified numerically (to machine precision) for all cases that we have
studied.
This removes the graviton weight of the ground state, which is always
absent for an odd number of electrons, because of angular momentum selection rules.

\begin{figure}[t]
\centering
{\includegraphics[height=2.8in,width=3.6in]{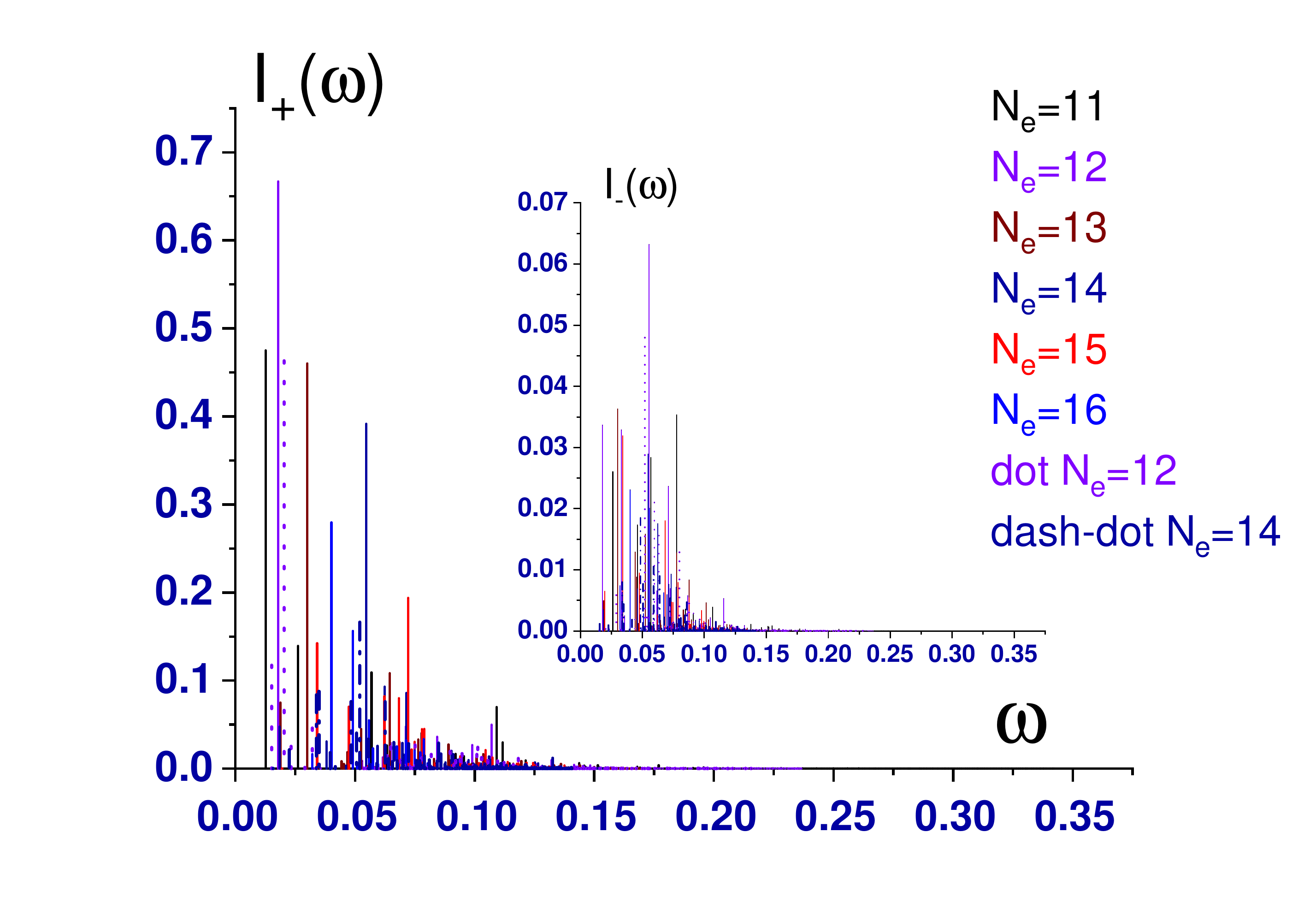}}
\caption{\label{fig:GS_Coul_Neg_3B}%
Color online -- Same as in Fig. \ref{fig:GS_Coul} except that we break PH symmetry by introducing a
3-body interaction potential, as described in the main text, with a strength of -0.01
The figure shows the spectrum for  positive
chirality.  The inset shows the spectrum for negative chirality, the response is seen to be
suppressed by an order of magnitude.
}
\end{figure}

\begin{figure}[b]
\centering
{\includegraphics[height=2.8in,width=3.6in]{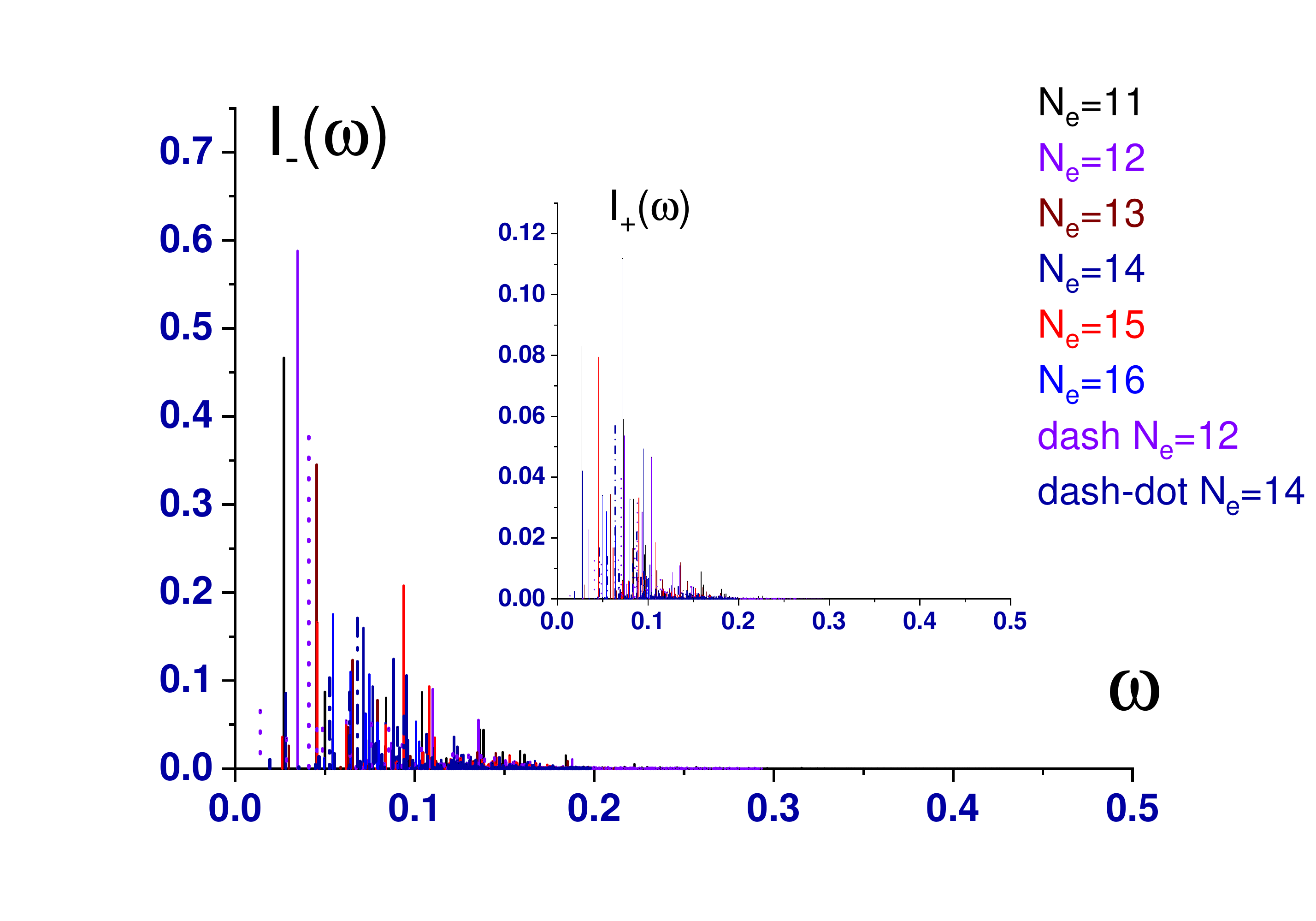}}
\caption{\label{fig:GS_Coul_Pos_3B}%
Color online -- Same as in Fig. \ref{fig:GS_Coul_Neg_3B} except we have added a 3-body pseudopotential with the opposite
sign (0.01). The stronger intensity is for  negative -2 chirality.  The inset gives
the same result but for positive +2 chirality, which is suppressed.
}
\end{figure}

We start with the case of pure Coulomb interaction. The PH symmetry is present in this case, and the ground state can be viewed as the PH-symmetrized MR state\cite{Rezayi2000}. As a result we have $I_{+}(\omega)=I_{-}(\omega)$, which are presented in Fig. \ref{fig:GS_Coul}. Similar to the cases studied in Ref. \cite{ChiralGraviton}, we observe fairly sharp peaks indicating the presence of graviton excitations in the system, except they come with both chiralities.
In Figs. \ref{fig:GS_Coul_Neg_3B} and \ref{fig:GS_Coul_Pos_3B} we show the graviton spectral functions in the presence of small 3-body PH symmetry breaking interactions. In calculating the
relative weights of two chiralities we normalize the weaker spectrum by the total weight of the
stronger.

Fig. \ref{fig:GS_Coul_Neg_3B} corresponds (roughly) to the case of LL mixing parameter $\kappa\approx 0.7$, which is representative of realistic situations, and tilts the ground state toward APf. While this results in a very small negative 3-body potential, it has a dramatic effect on the spectral functions: we find $I_+$ dominates $I_-$, with the total weight of the latter reduced to about $10\%$ of the former. This indicates gravitons with angular momentum $+2$ dominates the gravitational response of the system, which is in a hole-like APf state. In Fig. \ref{fig:GS_Coul_Pos_3B} we reverse the sign of the 3-body potential which favors the Pfaffian state, and the situation is reversed: $I_-$ dominates $I_+$, with the total weight of the latter reduced to about $20\%$ of the former. This indicates gravitons with angular momentum $-2$ dominates the gravitational response of the system, as we already saw in Ref. \cite{ChiralGraviton} for the Pfaffian state. We thus find the graviton chirality is opposite for the Pfaffian and APf states, and can be used to distinguish them experimentally (more on this point later).

{\em Hexagonal Geometry} --
Here, for an even number of electrons,
the topological sector is a set of 3-fold degenerate (related by rotations) ground states and
symmetry analysis of the graviton operator and the ground states is more complicated.
Notwithstanding, the
ground state weight can be removed by a set of
new orthogonal states, as was done above for the ZB doublets, except that the coefficients are
the cube roots of unity:
\begin{equation}
\lvert\Psi_0\rangle_a=\frac{\alpha\lvert \psi_0\rangle_1 +\beta\lvert \psi_0\rangle_2+\gamma
\lvert \psi_0\rangle_3}{\sqrt{3}}
\end{equation}
where $\alpha=e^{2i\pi/3}$, $\beta=e^{4i\pi/3}$, and
$\gamma = -\alpha-\beta$=1.
The other two states $\lvert\Psi_0\rangle_b$ and $\lvert\Psi_0\rangle_c$ are obtained by cyclic
permutaions of $\alpha$, $\beta$, and $\gamma$.
Again, the 3 expectation values of $\hat{O}^{(2)}_\pm$ adds to zero for all there ground state
as in the case for ZB doublets:
\begin{equation*}
\alpha ({_1\langle}\psi_0\lvert \hat{O}^{(2)}_\pm \rvert \psi_0 \rangle_1) +
\beta ({_2\langle}\psi_0\vert\hat{O}^{(2)}_\pm \rvert \psi_0 \rangle_2) +\gamma
({_3\langle}\vert\hat{O}^{(2)}_\pm \rvert \psi_0\rangle_3)=0.
\end{equation*}

\begin{figure}
\centering
{\includegraphics[height=2.8in,width=3.6in]{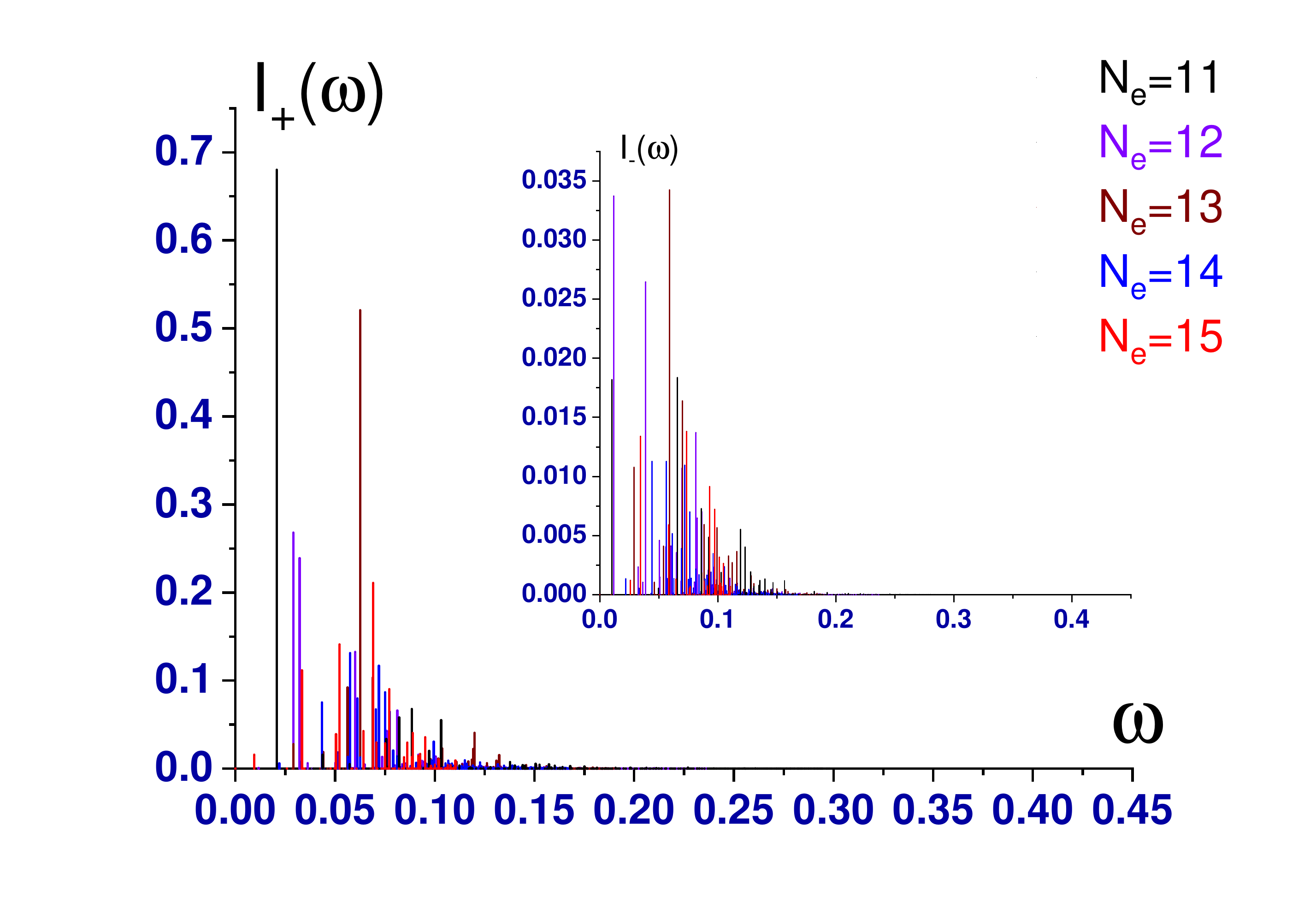}}
\caption{\label{fig:GS_Coul_Neg_3B_Hx}%
Color online -- Graviton spectral functions for 11-15 electrons on hexagonal geometry in the presence of a 3-body
potential of strength -0.01. As in the case of square geometry the +2 chirality is dominant
while the -2 chirality is strongly suppressed (inset).
}
\end{figure}

\begin{figure}[b]
\centering
{\includegraphics[height=2.8in,width=3.6in]{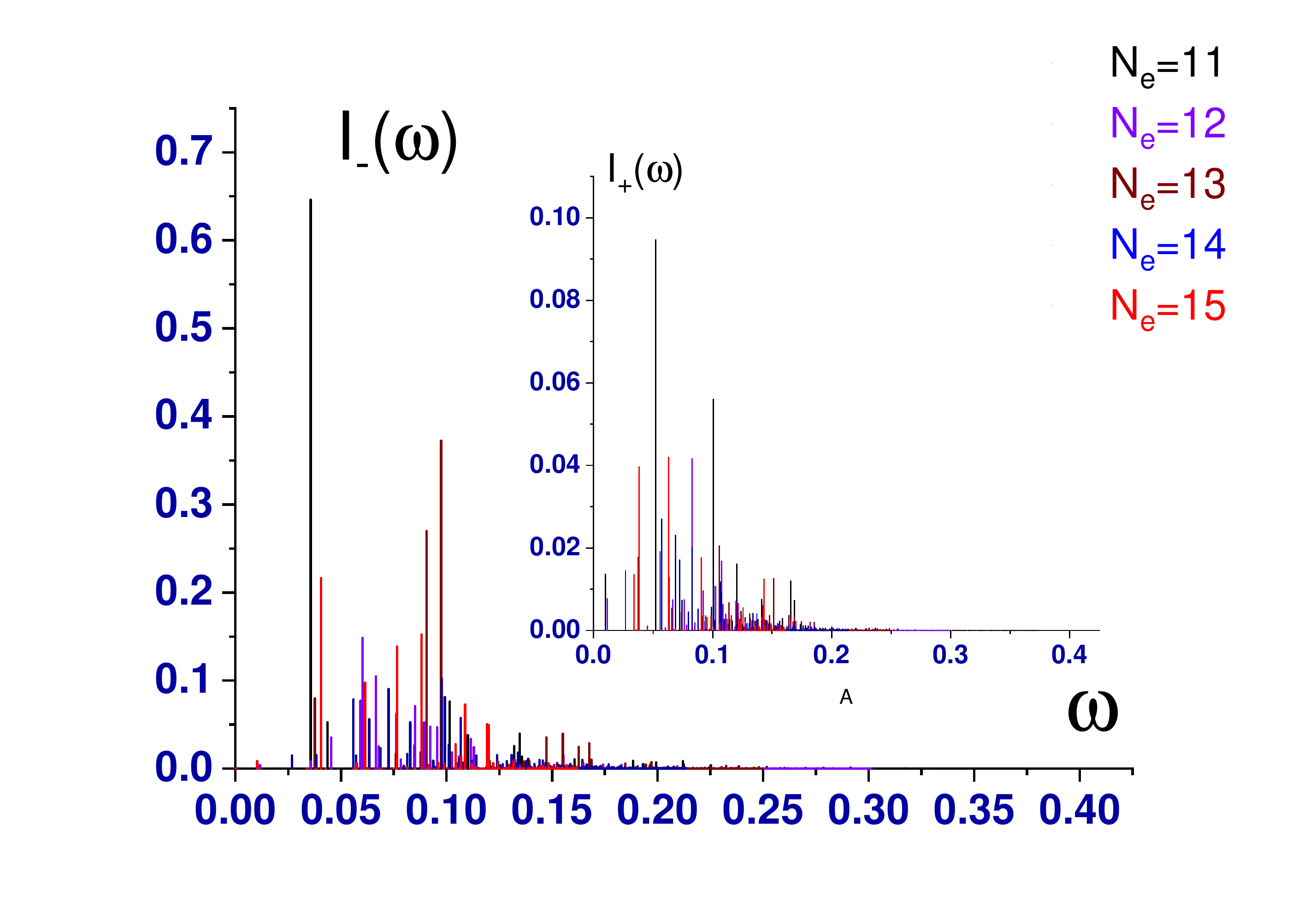}}
\caption{\label{fig:GS_Coul_Pos_3B_Hx}%
Color online -- Same as  Fig. \ref{fig:GS_Coul_Neg_3B_Hx} except for repulsive (0.01) 3-body potential.
The suppression of the opposite chirality (inset) is somewhat less suppressed than in Fig.
\ref{fig:GS_Coul_Neg_3B_Hx} chirality.  The inset shows the spectrum for positive chirality,
the response here is also suppressed.
}
\end{figure}

Figs. \ref{fig:GS_Coul_Neg_3B_Hx} and \ref{fig:GS_Coul_Pos_3B_Hx} are the hexagonal counterparts of Figs. \ref{fig:GS_Coul_Neg_3B} and \ref{fig:GS_Coul_Pos_3B}, where we see very similar behavior. The consistency between different geometries is indication that finite-size effects are minimal in our calculations.

We now return to the case of generic interactions for an odd number of particles.  The combination of anti-unitary
PH symmetry and discrete rotational symmetry could produce Wigner's extra degeneracies\cite{Haldane1985,papic_etal2012}.
The doublet appears for all sizes except when $N_e \mod(6)=1$.   In this case, the  PH partners
can each have a different angular momentum which leads to degeneracies. For the same reason
and conditions, the MR state and the Apf are orthogonal\cite{papic_etal2012}.
We find that our calculations already break the PH symmetry spontaneously.
However, because of the degeneracy
our code mixes the angular momenta of the doublet and as a result the matrix elements of the
graviton operator is ``contaminated'' and the weights become non-zero for every state.
The addition of a very small 3-body potential (of magnitude $-10^{-6}$) lifts the degeneracy and
restores the correct values of ground states angular momenta; the selection
rules reappear and the chiral graviton weights (many of them zero) look  very much like any other odd
electron case.

\begin{figure}
\centering
{\includegraphics[height=2.8in,width=3.6in]{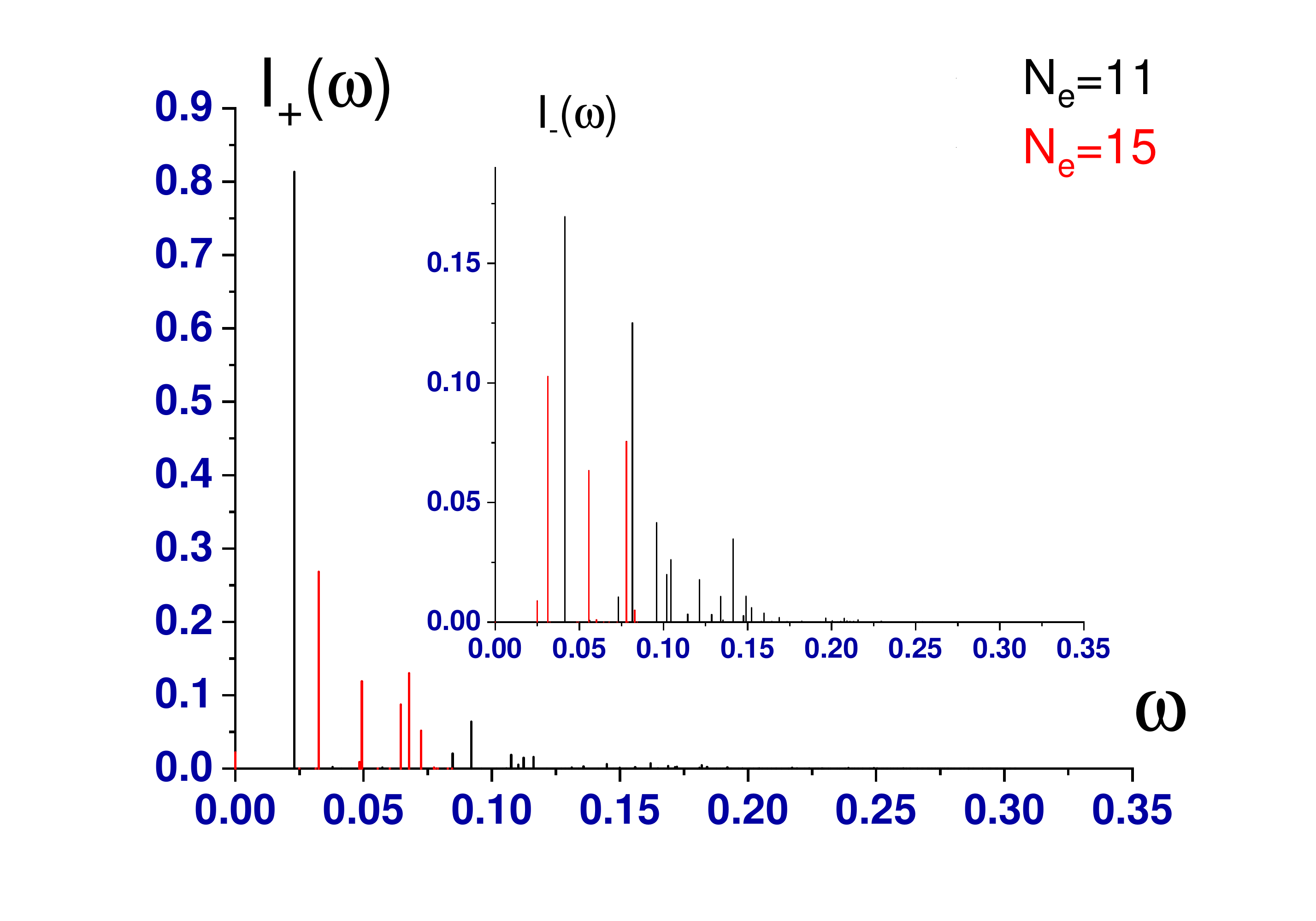}}
\caption{\label{fig:GS_OddN_Coul_small_3B}%
Color online -- The graviton spectrum for doublet ground states that can occur for odd
electrons (see main test). We have added a 3-body
 M=3 negative pseudopotential  $-10^{-6}$.  The system has broken PH
symmetry even before the 3-body interaction is added.  The red line at zero comes from the
doublet member which is split off by a very weak 3-body potential. The inset shows the spectrum for
negative chirality.  It is suppressed by a factor of $\approx 5.0$.  The added 3-body potential has
little rule in breaking PH symmetry.
}
\end{figure}

Fig. \ref{fig:GS_OddN_Coul_small_3B} shows that $I_+(\omega)$ is dominant while $I_-(\omega)$ is suppressed.  If the sign of
the 3-body pseudopotential is reversed then the plot looks the same, except $I_+$ and $I_-$ are exchanged.

{\em Discussion and Summary} --
We have calculated graviton spectral functions for Hamiltonians appropriate for the $\nu=5/2$ FQH state. While originally formulated as the system's response to a "gravitational wave"\cite{ChiralGraviton,yang}, it was anticipated that the gravitons and in particular their chiralities are detectable experimentally by
Raman scattering of circularly polarized light\cite{golkar,ChiralGraviton}.
In a very recent paper\cite{nguyen2021probing}, Nguyen and Son demonstrated that the Raman spectral functions are {\em identical} to the graviton spectral functions calculated here and in Ref. \cite{ChiralGraviton} (if the small anisotropy of the valence band is neglected), thus facilitating direct and quantitative comparison between theory and experiment. We note Pinczuck and coworkers' earlier results on the 1/3 Laughlin state\cite{Pinczuk} are in good agreement with our calculations\cite{ChiralGraviton}, although the graviton chirality could not be extracted since they used unpolarized light.

In sharp contrast to the 1/3 state, the situation is much murkier at 5/2, with many competing theoretical proposals. We demonstrated the leading candidates based on numerics, Moore-Read Pfaffian and anti-Pfaffian, can be clearly distinguished by the chiralities ($\mp 2$ respectively) of their graviton excitations, which are detectable using circularly polarized Raman scattering. We emphasize this is a {\em bulk} probe which does {\em not} suffer from many complications and subtleties at the edge. We note recent thermal transport experiments favor a particle-hole symmetric state at 5/2\cite{Banerjee2018,dutta2021novel}. This could be due to the presence of domains of Pfaffian and anti-Pfaffian states in the system\cite{Wan2016,Mross2018,Chong2018,Lian2018,zhu2020topological}. Such domains can also be revealed by Raman scattering, as long as their sizes are larger than the spatial resolution of the experiment. While we do not have a microscopic model that stabilizes an intrinsically particle-hole symmetric state, as discussed earlier we expect on general grounds that gravitons with both chiralities should be present and contribute (roughly equally) to the Raman scattering intensity of light with both circular polarization. We thus conclude polarized Raman scattering can potentially resolve all of the leading candidates for the 5/2 state.

\begin{acknowledgments}
We thank D. Nguyen and D. Son for useful conversations.
This work was supported by DOE grant No. \protect{DE-SC0002140}. KY's work was performed at the National High Magnetic Field Laboratory, which is supported by National Science Foundation Cooperative Agreement No. DMR-1644779, and the State of Florida.
\end{acknowledgments}
\bibliography{fqh_papers}
\end{document}